\documentclass[pdflatex,sn-mathphys-num]{sn-jnl}
\DeclareUnicodeCharacter{2043}{-}

\usepackage{graphicx}%
\usepackage{multirow}%
\usepackage{amsmath,amssymb,amsfonts}%
\usepackage{amsthm}%
\usepackage{mathrsfs}%
\usepackage[title]{appendix}%
\usepackage{xcolor}%
\usepackage{textcomp}%
\usepackage{manyfoot}%
\usepackage{booktabs}%
\usepackage{algorithm}%
\usepackage{algorithmicx}%
\usepackage{algpseudocode}%
\usepackage{listings}%

\usepackage{graphicx}
\usepackage{subcaption}   

\theoremstyle{thmstyleone}%
\theoremstyle{thmstyletwo}%

\theoremstyle{thmstylethree}%

\usepackage[T1]{fontenc}
\usepackage{lmodern}

\begin{document}

\title[Article Title]{Speaker-Disentangled Remote Speech Detection of Asthma and COPD Exacerbations}


\author*[1]{\fnm{Yuyang} \sur{Yan}}\email{yuyang.yan@maastrichtuniversity.nl}

\author[2,3]{\fnm{Sami O.} \sur{Simons}}

\author[1]{\fnm{Visara} \sur{Urovi}}

\affil*[1]{\orgdiv{Institute of Data Science}, \orgname{Maastricht University}, 
\orgaddress{\street{Paul-Henri Spaaklaan 1}, \city{Maastricht}, \postcode{6229 EN}, \country{the Netherlands}}}

\affil[2]{\orgdiv{Department of Respiratory Medicine, NUTRIM Research Institute of Nutrition and Translational Research in Metabolism, Faculty of Health Medicine and Life Sciences}, \orgname{Maastricht University}, \orgaddress{\street{P. Debyelaan 25}, \city{Maastricht}, \postcode{6229 HX}, \country{the Netherlands}}}

\affil[3]{\orgdiv{Department of Respiratory Medicine}, \orgname{Maastricht University Medical Centre}, \orgaddress{\street{P. Debyelaan 25}, \city{Maastricht}, \postcode{6229 HX}, \country{the Netherlands}}}


\abstract{

Early detection of exacerbations in asthma and chronic obstructive pulmonary disease (COPD) is important for timely intervention. Speech has emerged as a promising tool for continuous, non-invasive respiratory disease monitoring. However, speech signals inherently carry speaker-identifiable attributes that may dominate model predictions, which may compromise both diagnostic accuracy and patient privacy. We propose an adversarial learning architecture that disentangles pathology-related acoustic patterns from speaker-identifiable attributes through gradient reversal-based training. To enhance clinical interpretability, we employ SHapley Additive exPlanations (SHAP) to quantify the contributions of acoustic features to pathology-related predictions versus speaker identity. 
On the TACTICAS dataset, the adversarial model significantly reduces speaker separability, while improving or maintaining classification performance.
Specifically, it significantly improves the AUC for exacerbation type classification (from 0.65 to 0.73), enhances imbalance-sensitive metrics for respiratory status screening (PR-AUC and macro F1), and maintains cross-speaker generalization under the subject-wise split.
Concurrently, the J-ratio decreases, confirming effective suppression of speaker information. 
This study demonstrates that adversarial disentanglement of speaker identity not only enhances patient privacy but also maintains or improves diagnostic accuracy for respiratory disease monitoring. Our framework offers a foundation for deployable, interpretable, and privacy-aware speech-based methods in chronic respiratory disease management.

\medskip


\medskip

\textbf{Manuscript Statistics:} Total word count: 7215; Abstract word count: 193; Number of figures: 2; Number of tables: 5}

\maketitle

\section{Introduction}\label{sec1}

Asthma and chronic obstructive pulmonary disease (COPD) are among the most widespread chronic respiratory diseases, both characterized by persistent airflow limitation and frequently overlapping clinical presentations \cite{WHO}. Asthma and COPD lead to a main cause of global mortality, contributing to millions of fatalities yearly. Accurate and timely diagnosis is therefore important for appropriate interventions, reducing the risk of complications, hospitalizations, and healthcare costs.

Exacerbations are defined as sustained worsening of respiratory symptoms, which are the leading causes of emergency department visits and hospital admissions for COPD and asthma patients \cite{ghimire2021single, tsai2007factors}. Current clinical approaches to detecting exacerbations rely on patient-reported outcomes and standardized questionnaires (e.g., EXACT, CAT). However, these tools suffer from substantial variability between individuals due to subjective interpretation and variable symptom perception \cite{bhowmik2022personalized}. Moreover, distinguishing between asthma and COPD is important for guiding appropriate drug treatment. Misclassification is common in practice, studies indicate that over 50\% of COPD patients were initially mislabeled with asthma \cite{nissen2018concomitant}. Current guidelines recommend spirometry as the diagnostic foundation, complemented by comprehensive assessment of medical history, symptom patterns, and exposure risk factors \cite{vestbo2013global}.

While new techniques such as machine learning models utilizing computed tomography (CT) imaging \cite{moslemi2022differentiating} or electronic health records (EHRs) \cite{kocks2023diagnostic} have shown promise in respiratory disease detection, their clinical practice for continuous, real-world monitoring remains limited. Indeed, these existing approaches for respiratory disease assessment typically depend on diverse, clinic-based, or invasive inputs, providing only episodic clinical snapshots rather than longitudinal data. CT involves high costs and radiation exposure, making it impractical for frequent use. Similarly, EHR-based methods require structured clinical encounters and capture data only accessible during healthcare visits, posing barriers to accessibility. Consequently, these existing methods are fundamentally unsuited for continuous, real-world monitoring in patients' daily lives.
These limitations underscore the need for a low-burden, remote, and non-invasive modality capable of detecting respiratory disease outside traditional clinical environments.

Speech has emerged as a promising digital biomarker for respiratory health assessment \cite{xia2022exploring}. Clinicians noted perceptible vocal differences in patients during exacerbations \cite{van2021automatic}, reflecting the close connection between respiratory health and phonation. Voice production relies on coordinated respiratory airflow, a process that is disrupted by airway inflammation, dynamic hyperinflation, and altered breathing patterns in both asthma and COPD. Therefore, speech signals may encode subtle, objective features of underlying respiratory disease activity.

Recent studies have explored acoustic features from spectral, frequency, energy, and temporal domains, showing promising results for exacerbation detection \cite{yan2025developing}. 
Acoustic feature sets such as large-space extraction toolkit (openSMILE) and the extended Geneva Minimalistic Acoustic Descriptor Set (eGeMAPS) \cite{mayr2025assessing}, can effectively capture voice changes in COPD and asthma. Nevertheless, acoustic features often conflate pathology-relevant information with speaker-identifiable attributes (e.g., age, sex, language, or accent) \cite{dumpala2023manifestation}. Such entanglement poses two challenges: (1) it introduces spurious correlations that may degrade model generalizability across populations, and (2) it raises privacy concerns, as models may inadvertently memorize or leak sensitive demographic information.

Although methods such as sine-wave speech \cite{dumpala2021sine} or federated learning \cite{bn2022privacy} have been explored to enhance speech privacy, they often compromise diagnostic fidelity. Moreover, the extent to which acoustic features reflect respiratory pathology versus speaker-identifiable attributes remains poorly understood. To address these gaps, we propose a multi-task adversarial learning framework to disentangle pathology-related speech patterns from speaker-specific cues. By leveraging gradient reversal to suppress identity leakage while jointly optimizing for (i) respiratory status classification (stable vs. exacerbated) and (ii) exacerbation type classification (asthma exacerbation vs. COPD exacerbation), our approach seeks to simultaneously enhance diagnostic accuracy, and patient privacy.

The main contributions of this study are summarized as follows:

\(\bullet\)We provide a non-invasive, speech-based framework for remote respiratory disease monitoring in patients with asthma and COPD.

\(\bullet\)We propose a multi-task adversarial learning framework that simultaneously improves or maintains performance on two clinical tasks: (i) respiratory status classification (stable vs. exacerbated) and (ii) examining exacerbation type (asthma exacerbation vs. COPD exacerbation), while actively suppressing speaker-identifiable attributes.

\(\bullet\) We employ SHapley Additive exPlanations (SHAP) to provide model interpretability, revealing distinct sets of acoustic features associated with respiratory pathology versus speaker identity, thereby offering clinically meaningful insights into model behavior.

\(\bullet\) We validate the generalizability of our findings through external evaluation on the Bridge2AI-Voice dataset, confirming cross-dataset robustness.

To the best of our knowledge, this is the first study to address speaker-identifiable attributes in the context of speech-based respiratory disease monitoring through adversarial disentanglement. We validate the hypothesis that speaker-identifiable attributes in speech signals are irrelevant and often detrimental to respiratory disease monitoring. Their suppression simultaneously maintains or improves diagnostic accuracy and preserves privacy.

The rest of the paper is structured as follows: Section \ref{Sec:Background} describes the background of this work including COPD and asthma exacerbation, acoustic features and machine learning models. Section \ref{sec:Methods} outlines the dataset and techniques in this study. The detailed results of this work are presented in Section \ref{sec:Results}, followed a discussion of the findings in Section \ref{sec:Discussion}. Section \ref{sec:Conclusion} provides a conclusion of this work.

\section{Background}\label{Sec:Background}

\subsection{Exacerbations of COPD and asthma}
\label{sec:Exacerbations}

Exacerbations of COPD and asthma are characterized by an obvious worsening of respiratory symptoms such as dyspnea, cough, and sputum production, which last for several days to weeks and frequently require medical intervention or hospitalization. Cohort studies indicate that each exacerbation increases the risk of subsequent events and shortens the interval to the next exacerbation \cite{suissa2012long}. Consequently, early exacerbation detection and prompt initiation of treatment are important to mitigate disease progression and prevent complications.

Despite distinct underlying pathophysiologies, exacerbation of COPD and asthma often present with overlapping clinical presentations and diagnostic criteria, complicating differential diagnosis, particularly among populations with shared risk factors, such as current or former smokers and older adults \cite{kocks2023diagnostic}. Underdiagnosis or misclassification leads to increased emergency department visits, hospital admissions, mortality, and healthcare resource costs. 

To address these challenges, symptom-based clinical guidelines have been established by the Global Initiative for Chronic Obstructive Lung Disease (GOLD) and the Global Initiative for Asthma (GINA), providing standardized frameworks for diagnosis, prevention, and management. Nevertheless, real-world diagnostic accuracy remains suboptimal, more than 50\% of COPD patients were initially misclassified with asthma \cite{nissen2018concomitant}. 
Crucially, this misclassification often persists during exacerbations, leading to severe treatment errors. For example, antibiotics play a prominent role in the treatment of COPD exacerbations, however, antibiotics should not be used in primary care settings to treat asthma exacerbation unless there are clear signs of bacterial infection, as adjunctive antibiotic therapy has not been shown to reduce treatment failure or the risk of subsequent exacerbations \cite{mommers2025real}. Addressing this differential diagnosis specifically during exacerbations is therefore a critical clinical need. 

In response, data-driven methods have been explored to enhance clinical decision-making.
Zein et al. \cite{zein2021novel} built models from real-world EHRs data to predict asthma exacerbation. Lung functions \cite{wu2021acute} and breathing signals \cite{blouet2018prediction} were analyzed, showing remote monitoring of lung functions or breathing sounds may be used for the detection and/or follow-up of COPD exacerbation. 
Porter et al. \cite{porter2023smartphone} proposed a smartphone-based algorithm comprising cough analysis and patient-reported symptoms for asthma exacerbation detection. Additionally, speech-based approaches have demonstrated feasibility: Nippert et al. \cite{nippert2025detecting} used sustained vowel to detect COPD exacerbation before onset. Nallanthighal et al. \cite{nallanthighal2022detection} extracted acoustic features from speech recordings for COPD exacerbation detection.

However, many of these methods rely on additional clinical parameters such as forced expiratory volume in one second (FEV\textsubscript{1}), controlled breathing signals, or in-clinic recordings.
Because they require specialized sensors and equipment, controlled environments, or active patient effort, these methods are not practicable for continuous, real-world monitoring in daily clinical practice.
Cough-based systems also suffer from high inter-individual variability and dependence on patient effort, lacking standardized protocols. In contrast, spontaneous speech reflects natural vocal behavior, can be captured using ubiquitous mobile devices, without specialized sensors or clinical visits, offering a remote and non-invasive modality for respiratory disease monitoring.

\subsection{Acoustic features}
\label{sec:Acoustic biomarkers}

A wide range of acoustic features has been analyzed to capture respiratory changes in speech. Among these, spectral features such as harmonics-to-noise ratio (HNR), jitter, shimmer, Mel-frequency cepstral coefficients (MFCCs), are commonly used in speech analysis \cite{wkeglarz2025assessment}. MFCCs have the capacity to model vocal tract dynamics through the short-time power spectrum envelope, achieved an accuracy of 95.1\% for detecting lung disorders \cite{alghamdi2024deep}, highlighting the potential of acoustic biomarkers for rapid, non-invasive respiratory disease screening.

To overcome the limitations of purely spectral features, more comprehensive feature sets have been developed to capture a broader spectrum of vocal behavior. The extended Geneva Minimalistic Acoustic Parameter Set (eGeMAPS) \cite{eyben2015geneva}, for example, integrates low-level descriptors across different domains, has been successfully applied to distinguish post- versus pre-treatment status in COPD patients \cite{triantafyllopoulos2024sustained}.

In prior work, we demonstrated that optimizing the extraction parameters of MFCCs can enhance their performance in respiratory disease detection \cite{yan2025optimizing}. Building on this, we incorporated complementary features from the frequency, energy, and temporal domains to ensure a more comprehensive representation of voice signals. By fusing optimized MFCCs with these multi-domain acoustic features, we reported improved performance for distinguishing exacerbations in asthma and COPD \cite{yan2025developing}, laying the foundation for the feature set used in this study.

However, it is crucial to recognize that many of these acoustic features inherently carry speaker information. For instance, pitch is determined by the vibration and tension of vocal fold, varies with age and gender, making it a key feature in speaker recognition \cite{bai2004robust, carey1996robust}. Similarly, formant frequencies \cite{kent1993vocal, lariviere1975contributions} and MFCCs \cite{hosseinzadeh2007combining} model the vocal tract shape, which are highly unique to individual speakers and have been widely utilized in speaker identification systems.

\subsection{Machine learning models}
\label{sec:ML models}

Early approaches to respiratory disease detection from audio signals mainly employed conventional machine learning algorithms, including k-nearest neighbors (KNN), random forest (RF), support vector machines (SVM), and multilayer perceptrons (MLP). For instance, Venkata Srikanth et al. \cite{nallanthighal2022detection} applied an SVM classifier for COPD exacerbation detection, reporting an accuracy of 75\% and a sensitivity of 85\%. Alper et al. \cite{idrisoglu2024copdvd} compared RF, SVM, and CatBoost for COPD detection, achieved accuracies of 77 \%, 69 \%, and 78 \%, respectively. 

With the growing availability of audio datasets, deep learning has become as a powerful tool for acoustic modeling in respiratory health. Deep neural networks excel at enriching the representation of pathological voice by learning the complex, nonlinear mappings between input features and clinical labels. Among deep architectures, convolutional neural networks (CNNs) have been widely adopted for respiratory disease detection, typically operating on spectrogram inputs. The success of CNNs stems from their capacity to capture spatiotemporal patterns in time–frequency representations and translate them into discriminative physiological features. For example, Zeenat et al. \cite{9313208} employed CNNs with spectrograms derived from lung sounds to classify lung diseases, achieving highest accuracy of 95\% with data augmentation. Pre-trained models such as VGGish and BYOL-A have been utilized for continuous monitoring of respiratory quality of life \cite{despotovic2024digital}, highest AUC of 0.70 achieved by acoustic embeddings extracted with pre-trained BYOL-A model, demonstrating that vocal biomarkers can serve as possible alternatives for questionnaire-based clinical assessments. Nunavath et al. \cite{nunavath2018deep} applied an LSTM model and achieved an accuracy of 84.12\% to predict COPD exacerbation one-day ahead.

More recently, Transformer-based architectures have been explored for respiratory disease monitoring, including applications in COVID-19 detection and Asthma/COPD exacerbation management \cite{10342847, yan2025developing}. Transformers exhibit superior representational capacity for audio signals and have been shown to outperform CNNs in lung function estimation tasks due to their enhanced ability to model long-range dependencies, achieving the highest mean reciprocal rank of 0.655 for lung function estimation \cite{zhang2024towards}. Furthermore, advanced deep learning architectures, including attention-enhanced models and multimodal fusion, have demonstrated significant success in extracting robust patterns from noisy data streams in other complex domains, such as causal sentiment mining and web data analysis \cite{mihoubi2026attention, mihoubi2024discovering}.

\section{Methods}
\label{sec:Methods}

This study proposes an adversarial learning framework based on gradient reversal to disentangle speaker-identifiable attributes from pathology-related speech patterns, enabling privacy-preserving monitoring of exacerbations in asthma and COPD patients. Although prior analysis of the TACTICAS cohort demonstrated the feasibility of speech-based exacerbation detection \cite{yan2025developing}, it treated asthma and COPD patients as a single group and did not investigate the influence of speaker-identifiable attributes on model predictions. To support more precise clinical decision-making, we introduce a multi-task adversarial learning framework for remote respiratory disease monitoring.

The architecture optimizes two clinically hierarchical objectives: 

(i) respiratory status classification (stable vs. exacerbated), serving as an initial screening step to detect exacerbation onset.

(ii) exacerbation type differentiation (asthma exacerbation vs. COPD exacerbation), which acts as a differential diagnostic aid to guide disease-specific therapeutic strategies during the acute event. 

\begin{figure*}[h!]
   \centering
   \includegraphics[width=0.85 \columnwidth]{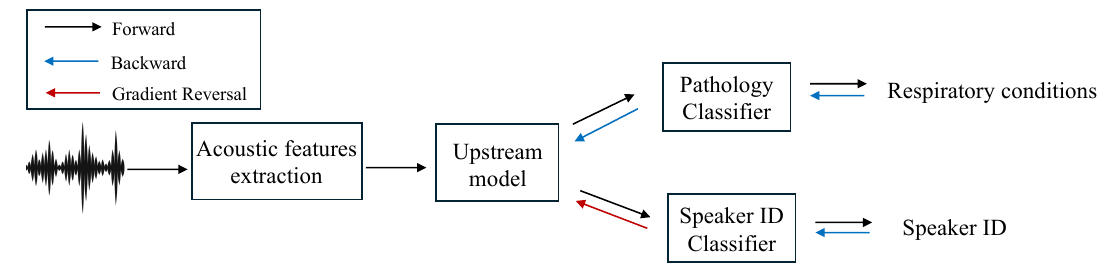}
  \caption{Our Proposed Methodology.}
  \label{fig:pipeline}
\end{figure*}

The overall architecture is depicted in Fig. \ref{fig:pipeline}, the shared upstream model minimizes the pathology classifier losses while simultaneously maximizing speaker classifier losses through a gradient reversal layer (GRL). This adversarial mechanism encourages the model to learn representations that are informative for respiratory pathology diagnosis yet invariant to speaker identity.

To systematically address our research questions, we designed three main experiments in this work:

(i) Speaker bias investigation (Section 4.1): We employed a state-of-the-art voice conversion technique (FreeVC) combined with a SVM model to investigate if the machine learning models rely on speaker-related attributes rather than pathology-relevant features. 

(ii) Adversarial Disentanglement (Sections 4.2 - 4.4): We trained the proposed multi-task adversarial learning framework to mitigate speaker dependency while optimizing for respiratory condition tasks.

(iii) External Validation (Section 4.5): We evaluated the generalizability and cross-lingual robustness of the proposed adversarial model on an external dataset (Bridge2AI-Voice).

\subsection{Dataset}
\label{Dataset}

This study uses data from the TACTICAS study \cite{TACTICASstudy, yan2025developing}, collected from a mobile application. Participants with diagnoses of asthma or COPD provided baseline demographic and clinical information during an initial hospital visit and subsequently submitted daily voice recordings over a three-month period. Each recording session included three tasks: sustained phonation of the vowel, a response to an open-ended question, and reading of a standardized passage. In this work, only spontaneous speech from the reading and question-answering tasks was analyzed, as they better reflect natural vocal behavior. Concurrently, participants completed the Exacerbations of Chronic Obstructive Pulmonary Disease Tool (EXACT) questionnaire every day, yielding severity scores ranging from 0 to 100. Respiratory states including exacerbation onset, recovery, and stable periods were determined according to the EXACT scoring criteria \cite{wedzicha2014extrafine} and further validated by a respiratory physician to ensure diagnostic fidelity. 


The final dataset supports two clinically hierarchical tasks: Task 1a performs respiratory status classification (stable as Class 0 vs. exacerbated as Class 1), serving as an initial screening step to identify exacerbations. Task 1b performs the same respiratory status classification (stable as Class 0 vs. exacerbated as Class 1), but in speaker-wise split. Upon detection of an exacerbation, Task 2 differentiates the underlying type (asthma exacerbation as Class 0 vs. COPD exacerbation as Class 1), thereby enabling appropriate therapeutic decisions.

To ensure robust evaluation, we applied tailored data splitting strategies for each task: 
Task 1a contains 8,704 recordings from 56 participants (7,900 stable and 804 exacerbated), we performed a stratified recording-wise split. Specifically, 20\% of the recordings were held out as the test set, while the remaining 80\% were allocated to training and validation, preserving the original class proportions across all subsets.
For Task 1b, we employed a speaker-wise split to guarantee that no single participant appeared in more than one subset. To construct this subset, we first excluded participants who provided recordings for only one respiratory state. For the remaining participants, who often exhibited severe class imbalance, the majority class was randomly downsampled to match the sample size of the minority class for that specific speaker. After balancing the number of recordings per speaker, this dataset consisted of 1,052 recordings (526 stable and 526 exacerbated) from 21 participants, 12 participants for training, 4 participants for validation, and 5 participants for testing. 
For Task 2, which focused exclusively on the 526 exacerbated recordings (312 COPD exacerbation and 214 asthma exacerbation) from the same 21 participants, we again adopted a subject-wise split to avoid speaker-level data contamination. In this case, 5 participants were designated as the test set, and the remaining 16 participants were divided into training (n=12) and validation (n=4) cohorts.



The Bridge2AI-Voice dataset \cite{b2ai_voice_2025} is used to evaluate the generalizability of the proposed adversarial model. This dataset comprises 19,271 voice recordings collected from 442 participants across five sites in North America, with all audio samples recorded in English. Participants were recruited based on conditions known to manifest in voice waveform including voice disorders, neurological disorders, mood disorders, and respiratory disorders. Among these, the dataset includes speech recordings from patients diagnosed with COPD (11 patients) and asthma (39 patients), with pre-extracted MFCCs features provided. To ensure class balance, we selected 11 asthma patients (234 recordings) and 11 COPD patients (249 recordings) for evaluation.

\subsection{Input Features}
\label{sec:Input Features}

We employed a fused acoustic feature set that integrates features from four complementary domains: spectral, frequency, energy, and temporal. This multi-domain representation was selected based on its demonstrated superiority over conventional handcrafted feature sets (e.g., eGeMAPS) and end-to-end deep embeddings in our prior speech analysis for respiratory disease monitoring \cite{yan2025developing}.

Specifically, the spectral features consist of the first 30 MFCCs, which capture vocal tract characteristics derived from the short-time power spectrum envelope. Complementary features from the frequency, energy, and temporal domains which are detailed in Table \ref{tab_features_set}, are integrated to capture a more comprehensive representation of speech signals. In the frequency domain, acoustic features such as pitch, jitter, and formant features were included, with their means and standard deviations calculated, resulting in a total of 12 features. The energy domain includes shimmer, loudness, and HNR, with 6 features extracted. Temporal features such as the rate of loudness peaks, the mean length of continuously voiced regions and unvoiced regions, and the number of continuous voiced regions per second, yielding 6 features in total.
This multi-domain integration avoids the risk of overlooking pathologically relevant features that may reside outside the spectral domain, thereby enhancing the model’s sensitivity to pathology-related speech alterations.

\begin{table}[h]
\caption{Acoustic features from multimodal domains.}\label{tab_features_set}%
\begin{tabular}{@{}ll@{}}
\toprule
Domain & Feature   \\
\midrule
Spectral & MFCCs (First 30 coefficients)  \\
\midrule
& Pitch (mean, std)   \\
Frequency & Jitter (mean, std)  \\
 & Formant frequencies (F1, F2, F3) (mean, std)  \\
 & Formant 1 (mean, std) \\
\midrule

 & Shimmer (mean, std)   \\
Energy & Loudness (mean, std)  \\
 & Harmonics-to-Noise Ratio (mean, std)  \\
\midrule

 & Rate of loudness peaks \\
Temporal & Duration of continuous voiced (mean, std)\\
& Number of continuous silence (mean, std)\\
 &  Voicing rate\\
\botrule
\end{tabular}
\end{table}

\subsection{Upstream Model}
\label{sec:Model - LeFF Transformer}

We adopted the Locally-enhanced Feed-Forward Network (LeFF) Transformer as the upstream feature encoder due to its demonstrated superiority over standard Transformer variants in modeling pathological speech patterns \cite{yan_interspeech2025}. Conventional Transformers employ a position-wise feed-forward network (FFN) that processes each token separately \cite{yuan2021incorporating}, thereby limiting their capacity to capture local dependencies among neighboring acoustic features. In contrast, the LeFF module was originally proposed for 2D image restoration, which enhances local contextual modeling by aggregating information from adjacent positions \cite{wang2022uformer}. We adapt this mechanism to the 1D acoustic domain, enabling the model to effectively capture short-range interactions across acoustic features from spectral, frequency, energy, and temporal domains.

The upstream architecture follows the design in our previous work \cite{yan_interspeech2025}. It begins with a linear projection that expands the input feature dimension, followed by a depthwise 1D convolution to capture localized contextual patterns. Non-linearity is introduced via the GELU activation function, and a dropout layer is applied to mitigate overfitting. A final linear layer projects the features back to the original channel dimension, ensuring compatibility with subsequent Transformer blocks. The upstream encoder consists of two LeFF Transformer blocks, each equipped with two self-attention heads. To further model sequential dynamics in speech, we append a Bidirectional Long Short-Term Memory (BiLSTM) layer with 128 hidden units.

\subsection{Adversarial Learning}
\label{sec:Adversarial Learning}

We employed an adversarial learning framework to disentangle pathology-relevant features from speaker-identifiable attributes. The upstream feature encoder generates shared embeddings from the input acoustic features. These embeddings are subsequently fed into two downstream task-specific classifiers: (i) a respiratory condition classifier, and (ii) a speaker identity classifier that serves as an adversarial discriminator.


The respiratory condition classifier is implemented as a three-layer multilayer perceptron (MLP) with ReLU activations and dropout (rate = 0.5) after each hidden layer. It maps the 64-dimensional upstream embedding to a logits vector corresponding to the target respiratory conditions. The speaker identity classifier shares the same MLP architecture. Inspired by the foundational domain-adversarial training framework \cite{ganin2016domain}, a Gradient Reversal Layer (GRL) is inserted between the upstream encoder and the speaker classifier. During forward propagation, the GRL acts as an identity function. During backpropagation, it multiplies the incoming gradients by a negative scaling factor (-$\lambda$), thereby reversing the gradient direction. This mechanism encourages the upstream model to learn representations that are informative for respiratory condition but uninformative for speaker identification. 

The total training objective is formulated as:

\begin{equation}
\mathcal{L}_{\text{total}} = \mathcal{L}_{\text{res}} - \lambda \mathcal{L}_{\text{spk}} 
\label{formula:L_total}
\end{equation}

Here, the negative sign encodes the adversarial nature of the optimization: it seeks to minimize the respiratory condition loss while simultaneously maximizing the speaker classification loss, thereby making the learned representations uninformative for speaker identification. The GRL serves as the computational mechanism to realize this objective during backpropagation.

Where $\mathcal{L}_{\text{res}}$ denotes the binary cross-entropy loss for the respiratory condition classifier:
\begin{equation}
\mathcal{L}_{\text{res}} = - \frac{1}{N} \sum_{i=1}^{N} [y_ilog(p_i)+(1-y_i)*log(1-p_i)]
\label{formula:L_EXA}
\end{equation}

Here, $N$ is the number of speakers in the training set. $y_i$ denoting the binary label indicating the respiratory condition of the $i^{th}$ speaker, $p_i$ is the model’s predicted probability for that speaker. The speaker identification loss $\mathcal{L}_{\text{spk}}$ is defined as the softmax cross-entropy, formulated as:
\begin{equation}
L_{spk} = - \frac{1}{N} \sum_{m=1}^{N} [log\frac{exp(x_{m,m})}{\sum_{j=1}^{N}exp(x_{m,j})}]
\label{formula:L_SPK}
\end{equation}

where $x_{m,m}$ represents the model’s output score assigning the $m^{th}$ speaker to its true speaker identity, and $x_{m,j}$ denotes the score for assigning the $m^{th}$ speaker to speaker $j$.

The hyperparameter $\lambda$ balances the trade-off between diagnostic accuracy and speaker invariance. 
To systematically select the optimal $\lambda$, we conducted a grid search over the candidate set \{1, 0.5, 0.1, 0.05, 0.01\}. Since this is a multi-objective optimization problem (aiming to maximize AUC while minimizing J-ratio), the final selection of the $\lambda$ was based on optimal performance of AUC and J-ratio on the validation set. 
This adversarial training strategy encourages the model to preserve or amplify pathology-related features for respiratory condition detection while suppressing speaker-identifiable attributes, thereby enhancing both diagnostic accuracy and patient privacy.

\subsection{Evaluation metrics}
\label{sec:Evaluation metrics}

Model performance is evaluated using recall, also known as the True Positive Rate (TPR) or Sensitivity, which quantifies the proportion of actual positive cases correctly identified by the model:

\begin{equation}
Recall = \frac{TP}{TP+FN}
\label{formula:Recall}
\end{equation}

where TP and FN denote the number of true positives and false negatives, respectively. Class-specific recall is reported for both classes in each binary classification task to account for potential label imbalance.

Given the class imbalance inherent in a certain task (Task 1a), relying on overall AUC and class-specific recall may provide an overly optimistic view of model performance. To provide a more rigorous and comprehensive evaluation, we additionally report three imbalance-sensitive metrics: Precision-Recall Area Under the Curve (PR-AUC), Balanced Accuracy, and Macro F1-score.

To assess the degree of speaker information retained in the learned representations, we compute the J-ratio \cite{guo2016speaker}, a measure of speaker separability based on within-class and between-class scatter matrices. The within-class scatter matrix $S_W$ and between-class scatter matrix $S_B$ are defined as:

\begin{equation}
S_W = \frac{1}{N} \sum_{i=1}^{N} R_i
\label{SW}
\end{equation}

\begin{equation}
S_B = \frac{1}{n} \sum_{i=1}^{N} (M_i - M_o)(M_i - M_o)^T
\label{SB}
\end{equation}

where  $N$ is the total number of speakers, $R_i$ is the covariance matrix of embeddings for the $i^{th}$ speaker, ${M}_i$ is the mean embedding vector for the $i^{th}$ speaker, and ${M}_0$ is the global mean embedding across all speakers. The J-ratio is then computed as:

\begin{equation}
J = trace[(S_W + S_B )^{-1} S_B ]
\label{jratio}
\end{equation}

A higher J-ratio indicates stronger speaker discrimination (i.e., more speaker identity leakage), whereas a lower J-ratio suggests effective disentanglement of speaker attributes from the learned representations. Embeddings extracted from the upstream LeFF Transformer model are used as input for J-ratio computation.

As a complementary and directly interpretable metric, we compute a speaker verification accuracy from the learned embeddings. Following the pairwise evaluation protocol commonly used in speaker verification \cite{brydinskyi2024comparison, desplanques2020tdnn}, the embeddings of every unordered pair of test-set recordings are compared using the cosine similarity, and a pair is predicted as originating from the same speaker when the similarity is equal to or greater than a decision threshold that is fixed on the validation set for each model. Speaker accuracy is then the proportion of pairs whose true same-speaker or different-speaker status is correctly predicted. A lower speaker accuracy indicates that less speaker dependency remains in the embeddings.

Point estimates are reported with two-sided 95\% confidence intervals obtained by bootstrapping with 5,000 resamples. Differences between the single-task baseline and the adversarial model were assessed with paired bootstrap tests, in which the same resample is evaluated by both models. 
A P value < 0.05 was considered statistically significant.

\subsection{Voice conversion}
\label{sec:Speaker_conversion}

Beyond privacy concerns, speaker-identifiable attributes embedded in acoustic features can lead to model overfitting to training speakers, thereby compromising generalization to unseen individuals. To investigate whether respiratory disease detection models rely on such speaker-specific cues rather than pathology-relevant features, a common approach is to use voice conversion to map all recordings to a single target speaker's voice before training the classifier \cite{ravi2024enhancing}. We conducted a voice conversion experiment using FreeVC \cite{li2023freevc}, a state-of-the-art text-free, one-shot voice conversion system. FreeVC disentangles speaker identity from linguistic and prosodic content by using self-supervised speech representations and an information bottleneck, enabling high-quality conversion without requiring textual annotations or parallel data, making FreeVC well suited for clinical speech dataset conversion such as TACTICAS. 
Mechanistically, FreeVC achieves high-quality conversion by disentangling linguistic content from speaker identity. It employs an information bottleneck to self-supervised speech representation (WavLM) features to extract source content information, while using a speaker encoder to extract and replace the speaker embedding. Consequently, FreeVC achieves higher F0-PCC (the Pearson correlation coefficient between F0 of source and converted speech), which demonstrates that FreeVC can effectively maintain the prosody of source speech.

To evaluate the impact of removing speaker identity on classification performance, we trained a SVM on the extracted acoustic features from the original and converted audios. The SVM serves as a classifier to clearly demonstrate the existence of speaker dependency by comparing the model performance on the original versus the voice-converted data.

\subsection{Model interpretation}
\label{sec:Model_interpretation}

To obtain insight into the mechanisms by which adversarial training improves diagnostic accuracy while suppressing speaker identity, we employed SHAP to quantify the contribution of each acoustic feature to model predictions. 
Specifically, for each feature we compute its mean absolute SHAP value across all N samples in the test set for both models. The plotted value represents the difference between the adversarial and baseline models: $\Delta \mathrm{SHAP} = |\mathrm{SHAP}|_{\text{Adversarial}} - |\mathrm{SHAP}|_{\text{Baseline}}$. A negative $\Delta$SHAP  indicates suppression, while a positive value indicates amplification.
This allows us to identify which features are most suppressed (i.e., their predictive contribution is reduced) and which are amplified (i.e., their contribution is enhanced) by the adversarial training process.

\section{Results}
\label{sec:Results}

We propose a multi-task adversarial learning architecture to disentangle pathology-relevant speech patterns from speaker-identifiable attributes, 
the results are described in following sections.


\subsection{Speaker Bias in Respiratory Condition Detection}
\label{ssec:Speaker-bias}

In this experiment, all speech recordings from the TACTICAS dataset were converted to a single target speaker’s voice with FreeVC. 
The acoustic features described in \ref{sec:Input Features} were extracted from the converted audios, and a SVM model was trained to evaluate performance on each task (Task 1a, Task 1b, and Task 2). 


As shown in Table \ref{tab_voice_conversion}, voice conversion results in a decline in classification performance. 
For Task 1a (stable vs. exacerbated, recording-wise split), the AUC decreases significantly from 0.89 to 0.81, with corresponding decreases in recall for the stable class (Class 0: 0.86 vs. 0.64), whereas recall for the exacerbation class (Class 1) remains comparable (0.77 vs. 0.78). 
For Task 1b (stable vs. exacerbated, speaker-wise split), the AUC drops significantly from 0.62 to 0.44, with a significant decrease in exacerbation recall (Class 1).
Similarly, for Task 2 (asthma exacerbation vs. COPD exacerbation), the AUC decreases significantly from 0.62 to 0.46, accompanied by significant decreases in stable class recall (Class 0).

Two complementary observations support the presence of speaker bias. First, the statistically significant degradation after voice conversion across all three tasks demonstrates that, the machine learning model exploits speaker-identifiable attributes as cues for classification, rather than learning robust, pathology-related acoustic biomarkers. Second, when evaluated on the original audio, the speaker-wise evaluation (Task 1b, AUC 0.62) yields lower performance than the recording-wise evaluation (Task 1a, AUC 0.89), indicating that overlapping speakers across splits may inflate apparent performance.

These findings confirm the presence of speaker bias, a limitation that undermines both clinical reliability and patient privacy, and underscore the necessity of speaker disentanglement mechanisms, which we address in Section \ref{ssec:Adversarial Learning}.

\begin{table}[htbp]
\centering
\caption{Impact of voice conversion (FreeVC) on respiratory condition classification performance.} 
\begin{tabular}{@{}llcccc@{}}
\toprule
\textbf{Task} & \textbf{Experiment} & \textbf{AUC} & \textbf{Recall (Class 1)} & \textbf{Recall (Class 0)} \\
\midrule
\multirow{2}{*}{Task 1a} 
    & SVM (Original Audio)     & 0.89 (0.87, 0.92) & 0.77 (0.71, 0.83) & 0.86 (0.85, 0.88) \\
    & SVM (FreeVC-converted Audio)   & \textbf{0.81$^{*}$ (0.76, 0.85)} & 0.78 (0.71, 0.84) & \textbf{0.64$^{*}$ (0.61, 0.66)} \\
\midrule
\multirow{2}{*}{Task 1b} 
    & SVM (Original Audio)     & 0.62 (0.55, 0.68) & 0.76 (0.69, 0.83) &  0.36 (0.28, 0.44) \\
    & SVM (FreeVC-converted Audio)   & \textbf{ 0.44$^{*}$ (0.37, 0.51)} & \textbf{ 0.53$^{*}$ (0.44, 0.61)} & 0.41 (0.33, 0.50) \\
\midrule

\multirow{2}{*}{Task 2} 
    & SVM (Original Audio)      & 0.62 (0.52, 0.71) & 0.75 (0.64, 0.85) & 0.46 (0.33, 0.59) \\
    & SVM (FreeVC-converted Audio)   & \textbf{0.46$^{*}$ (0.36, 0.56)}& 0.68 (0.57, 0.78) & \textbf{0.23$^{*}$ (0.13, 0.34)} \\
\bottomrule

\end{tabular}

\vspace{4pt}

\footnotesize  *, P value <0.05 when comparing performance between "Original Audio" and "FreeVC-converted Audio".

\label{tab_voice_conversion} 
\end{table}

\subsection{Adversarial Learning}
\label{ssec:Adversarial Learning}


Table \ref{table_adv_result} summarizes the performance of the single-task baseline and the proposed adversarial model on the TACTICAS dataset.
Both models share the same Transformer-based architecture described in Section 3, the baseline differs only in the absence of the speaker classifier and the Gradient Reversal Layer (GRL).
For Task 1a (stable vs. exacerbated, recording-wise split), the adversarial model achieves an AUC of 0.89, statistically comparable to the single-task baseline (0.90), while significantly improving recall for the stable class (Class 0: 0.84 vs. 0.81) and maintaining recall for the exacerbated class (Class 1: 0.81 vs. 0.80). Given the class imbalance of this task, the imbalance-sensitive metrics in Supplementary Table S1 are particularly informative: the adversarial model yields significant gains in PR-AUC (from 0.58 to 0.65) and macro F1 (from 0.66 to 0.69), together with a higher balanced accuracy. Given that the exacerbated class (Class 1) is the minority, the significant gains in PR-AUC and macro F1 indicate a more favorable precision–recall trade-off for the minority class, achieved at unchanged sensitivity and with significantly improved specificity (Class 0 recall).

For Task 1b (speaker-wise split), the adversarial model performs on par with the baseline across all metrics (AUC 0.60 vs. 0.60; balanced accuracy 0.60 vs. 0.59; macro F1 0.60 vs. 0.58; PR-AUC 0.57 vs. 0.58, none of the differences statistically significant), indicating that adversarial training does not yield classification gains under this split. Importantly, however, the speaker separability analysis (Section 4.3) shows that the adversarial model significantly suppresses speaker information in this setting, more details of speaker separability analysis will be discussed in following section.

For Task 2 (asthma exacerbation vs. COPD exacerbation),  the adversarial model significantly improves the AUC from 0.65 to 0.73, accompanied by a significant increase in recall for COPD exacerbation (from 0.53 to 0.68), while recall for asthma exacerbation remains unchanged (0.69 for both models). Consistently, balanced accuracy and macro F1 both improve significantly (from 0.61 to 0.68 and from 0.60 to 0.68, respectively), and PR-AUC increases from 0.74 to 0.79.


\begin{table}[htbp]
\centering
\caption{Performance comparison between the single-task baseline and the proposed adversarial model on the TACTICAS dataset.} 
\begin{tabular}{@{}llcccc@{}}
\toprule
\textbf{Task} & \textbf{Experiment} & \textbf{AUC} & \textbf{Recall (Class 1)} & \textbf{Recall (Class 0)} \\
\midrule
\multirow{2}{*}{Task 1a} 
    & Single-task baseline  & 0.90 (0.87, 0.92) & 0.80 (0.73, 0.86) &  0.81 (0.79, 0.83)\\
    & Adversarial (Ours)   & 0.89 (0.86, 0.92)& 0.81 (0.75, 0.86) & \textbf{ 0.84$^{*}$ (0.82, 0.86)} \\
\midrule

\multirow{2}{*}{Task 1b} 
    & Single-task baseline  &  0.60 (0.53, 0.67) &  0.66 (0.57, 0.74) & 0.51 (0.43, 0.60)  \\
    
    & Adversarial (Ours)   & 0.60 (0.53, 0.67) & 0.65 (0.57, 0.74) & 0.54 (0.46, 0.63) \\
\midrule

\multirow{2}{*}{Task 2} 
    & Single-task baseline  & 0.65 (0.56, 0.74) & 0.53 (0.40, 0.64) & 0.69 (0.57, 0.80) \\
    & Adversarial (Ours)
  & \textbf{0.73$^{*}$ (0.65, 0.82)} & \textbf{0.68$^{*}$ (0.57, 0.79)} &  0.69 (0.57, 0.80) \\
\bottomrule
\end{tabular}

\vspace{4pt}

\footnotesize "Single-task baseline" and "Adversarial (Ours)" models utilize the Transformer-based architecture described in Section 3. The baseline operates without the speaker classifier and Gradient Reversal Layer (GRL). *, P value <0.05 when comparing performance between "Single-task baseline" and "Adversarial (Ours)" models.

\label{table_adv_result} 
\end{table}

\subsection{Speaker Separability Analysis}
\label{ssec:SpeakerSeparabilityAnalysis}

To evaluate the extent to which speaker information is suppressed in the learned representations, we computed two complementary privacy metrics on the test-set embeddings: the J-ratio and the speaker verification accuracy. Lower values of either metric indicate that less speaker identity information remains in the embeddings.

As shown in Table \ref{spk_ratio}, adversarial training consistently reduces speaker separability across all tasks. 
For Task 1a (recording-wise split), the J-ratio decreases significantly from 1.54 in the single-task baseline to 1.03 in the adversarial model, and speaker accuracy drops significantly from 0.67 to 0.60, while the AUC remains statistically comparable (0.90 vs. 0.89).
For Task 1b (speaker-wise split), the J-ratio decreases significantly from 0.87 to 0.69, and speaker accuracy drops from 0.69 to 0.55, which is close to the chance level of 0.5, indicating that the adversarial embeddings are nearly uninformative about the identity of unseen speakers, while the AUC is unchanged (0.60 for both models).
For Task 2, the J-ratio decreases significantly from 1.00 to 0.80, accompanied by a small, non-significant reduction in speaker accuracy (0.75 vs. 0.73) and a significant improvement in AUC (from 0.65 to 0.73).

Importantly, this reduction in speaker dependency does not come at the expense of diagnostic accuracy: the AUC is preserved in Tasks 1a and 1b and significantly improved in Task 2. The concurrent suppression of speaker-identifiable attributes and preservation or improvement of classification performance demonstrates that removing speaker identity enables the model to focus more effectively on pathology-relevant acoustic biomarkers, thereby enhancing both clinical reliability and patient privacy.


To examine the sensitivity of the framework to $\lambda$, we evaluated AUC and J-ratio for all candidate values (Supplementary Table S2). The two objectives respond to $\lambda$ in opposite directions in that higher classification performance tends to exhibit weaker speaker suppression (higher J-ratio), reflecting the privacy–accuracy trade-off controlled by $\lambda$. This trade-off structure is stable across $\lambda$: the selected values reduce the J-ratio below the single-task baseline while maintaining or improving AUC, indicating that the conclusions of this study are not sensitive to the exact choice of $\lambda$, which instead determines the operating point on this trade-off.

\begin{table}[htbp]
\centering
\caption{J-ratio and AUC for the proposed adversarial model and single-task baseline.} 
\begin{tabular}{@{}llccc@{}}
\toprule
\textbf{Task} & \textbf{Experiment} & \textbf{J-ratio} & \textbf{Speaker Accuracy}& \textbf{AUC} \\
\midrule
\multirow{2}{*}{Task 1a} 
    & Single-task baseline  & 1.54 (1.44, 1.82)& 0.67 (0.65, 0.69) &  0.90 (0.87, 0.92)  \\
    & Adversarial (Ours)   & \textbf{1.03$^{*}$ (0.94, 1.28)}& \textbf{ 0.60$^{*}$  (0.58, 0.63)} & 0.89 (0.86, 0.92)   \\
\midrule

\multirow{2}{*}{Task 1b} 
    & Single-task baseline  & 0.87 (0.68, 1.17)&  0.69 (0.65, 0.72) & 0.60 (0.53, 0.67) \\
    & Adversarial (Ours)   & \textbf{ 0.69$^{*}$ (0.52, 0.97) }&  \textbf{0.55$^{*}$ (0.51, 0.59)} & 0.60 (0.53, 0.67)   \\
\midrule

\multirow{2}{*}{Task 2} 
    & Single-task baseline  &  1.00 (0.66, 1.71)& 0.75 (0.71, 0.77) & 0.65 (0.56, 0.74) \\
    & Adversarial (Ours)
  & \textbf{ 0.80$^{*}$ (0.62, 1.23) } & 0.73 (0.68, 0.77)  & \textbf{0.73$^{*}$ (0.65, 0.82)} \\
\bottomrule
\end{tabular}

\vspace{4pt}

\footnotesize  *, P value <0.05 when comparing performance between "Single-task baseline" and "Adversarial (Ours)" models.

\label{spk_ratio} 
\end{table}

\subsection{Suppressed Features Interpretation}
\label{ssec:SHAP}

\begin{figure}[h]
\centering

\begin{subfigure}{\textwidth}
\centering
\includegraphics[ height=6cm]{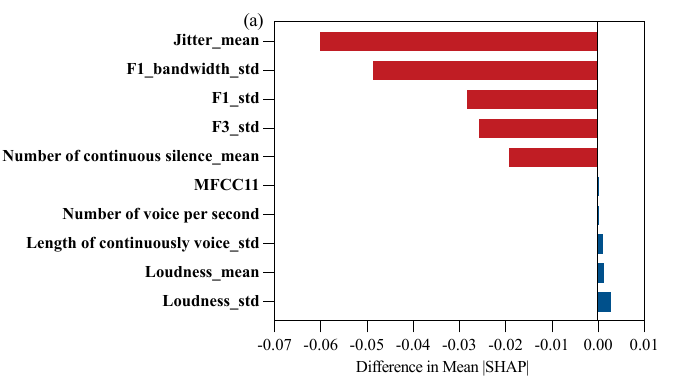} 
\label{fig:shap_1}
\end{subfigure}

\vspace{0.25cm}

\begin{subfigure}{\textwidth}
\centering
\includegraphics[height=6cm]{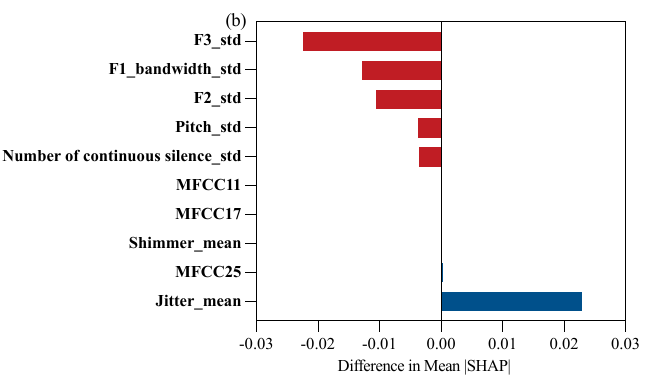}
\label{fig:shap_2}
\end{subfigure}

\caption{Top Features by Absolute Importance Change, (a) Task 1b, (b) Task 2.}
\label{fig:shap}
\end{figure}

Fig. \ref{fig:shap} presents bar plots of the features most suppressed (red) and most amplified (blue) by adversarial model, ranked by the absolute change in mean |SHAP| values between the single-task baseline and the adversarial model on the test set. Because Tasks 1b and 2 share the same participants and the same speaker-wise splitting strategy, we focus the interpretation on these two tasks, the corresponding analysis for Task 1a is provided in Supplementary Fig. S1.



For Task 1b (stable vs. exacerbation, speaker-wise split), the adversarial model suppresses features that are strongly correlated with speaker identity, such as jitter and F1 bandwidth, and the first and third formant frequencies (F1, F3). Formant frequencies encode vocal tract shape and jitter reflects the variability of F0, both vary little within a speaker and are widely used in speaker verification \cite{farrus2007jitter}, so they carry little information about within-speaker changes in respiratory state while reflecting more about speaker identity. Concurrently, the model amplifies features associated with disruptions in breathing-voice coordination, including loudness (mean and std), the variability of continuously voiced segment lengths, and the number of voiced segments per second. This shift from speaker traits to state-dependent respiratory features is consistent with the reduced J-ratio and speaker accuracy reported in Section 4.3.


For Task 2 (asthma exacerbation vs. COPD exacerbation), a distinct pattern emerges. 
Speaker-related features such as pitch, F2, F3, and F1 bandwidth are again suppressed, as they offer no discriminative value between the two diseases. In contrast, jitter is amplified. Because Task 2 discriminates between diseases rather than states, voice-perturbation characteristics become informative pathology markers: COPD, with fixed airway obstruction and parenchymal damage, produces more severe and persistent vocal perturbations than asthma, whose structural changes are largely reversible \cite{saeed2018study}. The model therefore leverages disease-specific perturbation features rather than speaker identity or exacerbation markers.

The corresponding analysis for Task 1a (recording-wise split, Supplementary Fig. S1) shows a partially opposite pattern, with jitter and F1 bandwidth amplified rather than suppressed. When the same speakers appear in both training and test sets, the variability of F0 can be exploited as speaker-specific reference cues, inflating their importance. This is consistent with the speaker dependency demonstrated in Section 4.1, where performance under the recording-wise split is markedly higher than under the speaker-wise split and degrades after voice conversion. The suppression of these same features in Task 1b illustrates that adversarial training removes speaker-confounded cues that do not transfer to unseen speakers.

A more detailed interpretation of the physiological and clinical relevance of these acoustic features will be provided in the following Discussion section.

\subsection{External Dataset Validation}
\label{ssec:validation}


The architecture of both the baseline and adversarial model for external dataset validation remained consistent with those described in Section \ref{sec:Methods}. To evaluate cross-dataset generalizability, both models were applied to the Bridge2AI-Voice dataset.
It is important to clarify that the labels in the Bridge2AI-Voice dataset represent baseline disease diagnosis (asthma vs. COPD) and do not contain annotations for exacerbation status. Due to the extreme scarcity of publicly available speech datasets with annotated exacerbation states, Bridge2AI-Voice serves as the best available speech data for external validation. Moreover, it provides a test for the generalization of Task 2 (differentiating asthma from COPD). Specifically, it evaluates whether the disentangled, pathology-relevant features learned from the Dutch TACTICAS cohort can generalize to differentiate the underlying diseases in a completely different cohort, and language (English).

As shown in Table \ref{table:validation}, the adversarial model achieves an AUC of 0.92, significantly outperforming the single-task baseline, with a significant improvement in recall for the COPD class and comparable recall for the asthma class. Consistently, the adversarial model yields significant gains in macro F1 (0.78 vs. 0.70) and PR-AUC (0.96 vs. 0.90), together with a higher balanced accuracy (0.80 vs. 0.74, Supplementary Table S3).
The J-ratio decreases from 2.09 to 1.59, and speaker accuracy remains comparable, neither difference reaches statistical significance, likely reflecting the limited size of the external cohort (22 participants), however, the direction of the J-ratio change is consistent with the speaker suppression observed on TACTICAS.

These results demonstrate that the proposed framework maintains its diagnostic advantage when applied to an external, English-language dataset, confirming its robustness and cross-dataset, cross-lingual generalizability, while the privacy benefit remains directionally consistent with the TACTICAS analyses.



\begin{table}[ht]
\centering
\caption{Bridge2AI-Voice dataset validation. }
\begin{tabular}{ c  | c  | c | c | c}
\hline
& AUC & Recall-COPD  & Recall-asthma  & J-ratio  \\
\hline
Single-task Baseline  &  0.79 (0.70, 0.87)  & 0.69 (0.59, 0.78)& 0.79 (0.66, 0.92) &  2.09 (1.54, 3.08) \\
\hline
Adversarial (Ours) & \textbf{ 0.92$^{*}$ (0.87, 0.97)   }  & \textbf{ 0.78$^{*}$ (0.70, 0.87) }   &  0.82 (0.68, 0.92)  & 1.59 (1.11, 2.61)\\
\hline

\end{tabular}

\vspace{4pt}

\footnotesize  *, P value <0.05 when comparing performance between "Single-task baseline" and "Adversarial (Ours)" models.

\label{table:validation}
\end{table}

\section{Discussion}
\label{sec:Discussion}

This study presents a multi-task adversarial learning framework that simultaneously addresses two challenges in speech-based respiratory disease monitoring: diagnostic accuracy and patient privacy. By disentangling pathology-relevant acoustic patterns from speaker-identifiable attributes, our approach not only improves or maintains classification performance across the clinically hierarchical tasks: (i) respiratory status classification (stable vs. exacerbated) and (ii) exacerbation type classification (asthma exacerbation vs. COPD exacerbation), but also reduces speaker dependency that compromises generalizability and raises privacy concerns. 

The clinical motivation for this hierarchical design aligns with established respiratory care pathways. Early detection of exacerbations is important to prevent disease progression, reduce symptom severity, and avoid hospitalization. Subsequent disease differentiation between asthma and COPD further enables appropriate pharmacological management. 
Our results validate this workflow: for the recording-wise screening Task 1a, the adversarial model achieves statistically comparable AUC to the baseline, while significantly improving specificity, PR-AUC, and macro F1. 
For Task 2, it also significantly outperforms the baseline. 
Importantly, the subject-wise evaluation (Task 1b) and the voice conversion experiment reveal that, when speakers overlap across splits, part of the apparent performance under the recording-wise split derives from speaker-specific cues: cross-speaker AUC drops to 0.60, and removing speaker identity via FreeVC significantly degrades performance in all tasks. These findings confirm that speaker-identifiable attributes act as confounders in speech-based respiratory monitoring and justify the disentanglement strategy.

The necessity of such disentanglement is demonstrated by the FreeVC voice conversion experiment: when all speakers are converted to a single target speaker, classification performance declines significantly in all tasks, confirming that baseline models exploit speaker-specific attributes as proxy features. 
In contrast, the adversarial framework significantly reduces speaker separability (J-ratio and speaker accuracy) on TACTICAS while preserving or improving diagnostic accuracy, and demonstrates robust cross-dataset generalizability: on the external, English-language Bridge2AI-Voice dataset, the adversarial model significantly improves the AUC, PR-AUC, and macro F1. The reduction in J-ratio on the external dataset was directionally consistent with the TACTICAS results but did not reach statistical significance, likely owing to the small external cohort. 


The mechanism underlying performance enhancement is further illuminated by the SHAP-based interpretability analysis. 
In Task 1b, the adversarial model suppresses features strongly associated with speaker identity. These features include formant frequencies (F1, F3), which encode vocal tract shape and are widely used in speaker verification systems \cite{kent1993vocal}, together with jitter and F1 bandwidth. Jitter fundamentally reflect individual glottal configurations and the inherent stability of vocal fold vibration, and research has shown that speaker identity is remarkably robust and well-preserved even under changes in formant bandwidths \cite{kuwabara1987role}. Consequently, they act as confounding identity features rather than indicative markers of respiratory changes associated with exacerbations.
Concurrently, the model amplifies features linked to breathing–voice coordination disorders, including loudness, for which higher values have been reported in healthier voice compared with dysphonic or COPD voice \cite{wkeglarz2025assessment}, and temporal features such as the variability of continuously voiced segment lengths and the voicing rate, reflecting the altered breath-pausing patterns during exacerbation \cite{wiechern2018effects}. This shift from speaker traits to pathology-related features is consistent with the reduced J-ratio and speaker accuracy, and explains how the adversarial model preserves cross-speaker generalization.

For Task 2, a distinct yet complementary pattern emerges: speaker-related features such as pitch, which is determined by vocal-fold vibration and varies systematically with age, sex, and body size \cite{bai2004robust}, formant frequencies (F2, F3) and F1 bandwidth are again suppressed, as they offer no discriminative value between the two diseases. Instead, jitter and shimmer are selectively amplified, suggesting that the degree and nature of vocal-fold vibration irregularity differ between the two conditions. This aligns with reported findings in \cite{saeed2018study}, COPD typically involves fixed, progressive airway obstruction and parenchyma (e.g., emphysema), leading to more severe and persistent vocal perturbations, whereas asthma structural changes are largely reversible. 
Together, these shifts confirm that the adversarial model prioritizes clinically meaningful features over speaker fingerprints, thereby improving robustness and interpretability.

Nevertheless, several limitations and challenges are associated with this approach. First, while the voice conversion (FreeVC) experiment provides evidence of speaker bias, it is inherently an indirect measure. State-of-the-art voice conversion systems may introduce acoustic artifacts alongside speaker identity. Future work should explore more pathology-preserving disentanglement methods or utilize synthetically controlled datasets where speaker identity and pathological states can be manipulated independently to provide more rigorous causal evidence.
Second, although external validation was performed, the Bridge2AI-Voice dataset provides only disease labels (asthma vs. COPD) and contains no recordings annotated for exacerbation. Future work should evaluate the framework across more diverse datasets that also capture exacerbation states. Third, the current study focuses exclusively on asthma and COPD, inclusion of more respiratory conditions (e.g., bronchiectasis) would strengthen the framework’s differential diagnostic utility.
Finally, the subject-wise evaluations are based on a limited number of participants (21 speakers), and the modest performance of Task 1b indicates that generalizability to unseen speakers remains challenging and will require larger multi-speaker cohorts.


In summary, this work establishes that privacy-preserving adversarial learning is not only an ethical safeguard but also a technical promoter of clinical fidelity. By reconciling diagnostic accuracy with speaker invariance, our framework advances the vision of equitable, interpretable, and deployable speech-based digital biomarkers for chronic respiratory disease management.

\section{Conclusion}
\label{sec:Conclusion}

This study presents an adversarial learning framework designed to enable accurate, privacy-preserving remote monitoring in patients with asthma and COPD. By disentangling pathology-relevant acoustic features from speaker-identifiable attributes, our approach simultaneously maintains or enhances diagnostic accuracy and mitigates privacy risks. Interpretability analysis further reveals that the model suppresses speaker-specific features while amplifying pathology-relevant biomarkers. These findings provide an explanation for the observed performance improvements and declare the success of the adversarial disentanglement strategy. Moreover, external validation on the Bridge2AI-Voice dataset confirms the model’s cross-dataset and cross-lingual generalizability. This work thus provides a robust foundation for the future deployment of equitable, interpretable, and clinically actionable respiratory monitoring systems in real-world settings, where continuous, non-invasive, and privacy-conscious assessment is essential for timely intervention and personalized care.

\section*{CRediT authorship contribution statement}


Yuyang Yan: Methodology, Conceptualization, Writing – Original Draft, Investigation. Sami O. Simons: Writing – Review \& Editing, Supervision. Visara Urovi: Writing – Review \& Editing, Supervision.

\section*{Acknowledgments}

The authors would like to thank Loes van Bemmel for the collection of the TACTICAS dataset, and Julia Hoxha from Zana Technologies GmbH who provided the mobile application for TACTICAS data collection.

\section*{Declaration of competing interest}
The authors declare that they have no known competing financial interests or personal relationships that could have appeared to influence the work reported in this paper.

\section*{Ethical approval}
This study was approved by the Ethical Review Committee Inner City Faculties (ERCIC), with reference number: ERCIC\_528\_31\_01\_2024





\bibliography{sn-bibliography}

\end{document}